# How to Train your Space Tester:
# Big Picture Challenges Facing Space Vehicle Test

Michael Nayak[1], Christina Straight[2], Evelyn Kent[3] and Jarred Langhals[4].

1 (**M**): 416th Flight Test Squadron, Air Force Test Center; michael.nayak.1@us.af.mil
2: Air Force Research Laboratory, Space Vehicles Directorate, christina.straight@us.af.mil
3: Air Force Research Laboratory, Space Vehicles Directorate, evelyn.kent@us.af.mil
4: Innovation and Prototyping Directorate, Space and Missile Systems Center, jarred.langhals.2@spaceforce.mil

**Abstract**: To ultimately benefit the space warfighter, test fundamentals and test conduct under stresses of time and fuel will be where the rubber meets the road. Fundamental big-picture challenges face the enterprise of burgeoning space test, which may also have an impact on the syllabus and training of a possible future "Space Test Pilot School". This paper specifically discusses four of those challenges, and their impact on US Space Force operators:
  (1) Testing limits of space capabilities: Space vehicles have traditionally imposed very strict operating limits. A culture shift from the top-down will be required before space testers are allowed to test spacecraft limits as freely as air-breathing test pilots do today.
  (2) Permitted procedures: There are a limited number of contingency procedures at a space operator's disposal, further constrained within themselves.
  (3) Safety nets: A discussion of space-specific methods that mitigate risks in operations, and the need for more agile fault-detection.
  (4) Operator proficiency: The qualifications and recency of a tester at the "controls" must be considered in the context of a single misstep potentially resulting in the entire asset being lost. We suggest that *hours in high-risk test per spacecraft* may become the defining factor for high-risk space test execution.

## 1. Introduction

For 50 years, the United States has enjoyed an unparalleled dominion of space, including relative freedom to conduct sensitive operations by leveraging the vastness of the space domain. As noted by [*Nayak*, 2018], this is not today's reality; the focus of near-peer rivals is on a critical susceptibility of our war-making enterprise: the near-ubiquitous use of space power [*Hucheng*, 2000; *Nayak*, 2016; *Stewart*, 2016]. The formation of US Space Force (USSF) is a response to the contested, degraded, and operationally limited environment of space today.

Numerous recent space policy essays [e.g. *Anthony*, 2019; *Cox*, 2019] discuss the warfighting mindset that the burgeoning USSF and its space operators needs to embody in order to be effective in its mission. However, operations begin with test. To eke all the capabilities out of a space vehicle, and determine operational limits for warfighter employment of spacecraft, USSF needs testers at all levels with a considered risk posture who can push vehicles to their limits in a safe and rigorous manner.

The venerable US Air Force Test Pilot School (USAFTPS) is among the gold-standard institutions for training pilots, navigators and engineers to evaluate and test complex aerospace systems. Discussions are currently ongoing between USSF and USAFTPS to develop and propagate a space test education program similar to the twelve-month "long course" that TPS graduates receive, which would have far-reaching consequences for both institutions. For now, USAFTPS is hosting a 12-week Space Test Fundamentals Course to increase the cadre of testers within USSF. While classroom education is certainly a step in the right direction, to ultimately benefit the warfighter, test fundamentals and test conduct under stresses of time and fuel will be where the rubber meets the road. It is our belief that there are some fundamental big-picture challenges facing the field of space test which need to be addressed first, and may have an impact on the syllabus and training of a forthcoming Space Test cadre.

The authors of this work have all logged time as Flight Directors or Lead Flight Directors of different high-risk experimental (prototype) space vehicles, to include operations and test with AFRL/RV, the research laboratory organization responsible for developing/testing "experimental" satellites. One author has additional experience with F-16 air test, and is a USAFTPS graduate. All authors have tested and transitioned these vehicles to operational Space Operations Squadrons (SOPS) within what is today USSF (formerly Air Force Space Command). From this experience, we can confidently say that there is a large gap between the mentality of air and space test. In the air domain, on a tactical level, it is well-accepted that a well-trained four-ship acting independently can exploit a central-node directed denial system. This is accomplished by getting inside the OODA loop of an enemy that operates as a geographically-separated central command giving operators sitting at the trigger the authorization to fire. Today, we posit that while US air operations have evolved toward more refined and lethal independent four-ship CONOPS, US space operations remains stagnant in the old Soviet tactical air paradigm [*Drane*, 1976], where operators do not have independent authorization to employ their vehicle as they see fit in a warfighting domain. In most cases, space operators are not authorized to conduct a recovery of the vehicle from a fault state without executive-level approval, unless it is a known and recurring fault. Across this paper, we will discuss some core differences between the two domains, and considerations for space test that may have implications for the developing curriculum to be taught to future Space TPS-trained professionals, as well as space test as an enterprise whole.

## 2. Finding the Edges of the Envelope: Testing the Limits of Space Vehicle Capabilities

The quest to find and define the "corners" of the performance envelope is an endeavor fundamental to airplane test. Starting from a safe envelope for initial operation where the flight dynamics are predictable and constant (heart of the envelope), and expanding the flight clearance to an operational envelope that can be flown with "carefree handling" by any mission-qualified pilot operator, is a core tenet of a new aircraft's flight test program. This campaign is not without acknowledged risk; it is impossible to find or expand the limits of a system under test (SUT) without acknowledging that there is an inherent risk of losing or damaging it. The risk of losing the system should not and does not stop testing in its tracks; rather, flight testers are trained to manage and mitigate these risks to ensure a successful outcome [*Lowry, 2012*]. Ultimately, the edges of the envelope are discovered, documented and then adhered to by use of operating restrictions and flight manual instructions. In summary: a SUT's capabilities cannot truly be known unless it is tested at conditions close to its design limits, which drives the selection and execution of envelope expansion test points.

In the space domain, even if a vehicle is not one-of-a-kind (e.g., a GPS satellite), there are typically enough new and unique features on the spacecraft to make it equivalent to a test asset. In addition to this, the difficulty of obtaining access to orbit, and near-impossibility of mid-lifespan space vehicle repair, required a leadership and operator mindset that the space vehicle must be protected at all costs. At AFRL/RV, test/experimentation operators and planners are given more leeway in risk acceptance, as such spacecraft are developed to be experimental. But especially for operational missions, the community places an extremely high value on space assets, to the point where they are unwilling to risk any damage to a vehicle. The cost of a single spacecraft, in money, in strategic (single-string) capability loss, and time, further adds to this intractability in terms of risk.

A few notes about the illusion of costs and technology lead time is worthwhile here, as both can be overcome. AFRL matures space-bound technology by flight testing critical new technologies, techniques and Concepts of Operation, which offsets the predominant non-recurring engineering costs [*Sellers et al.*, 2015] while lowering technical risk for use in future systems. DoD and others have turned their focus to open systems architectures in acquisitions, providing flexibility to procure components from a wide variety of vendors with little forethought. This also allows for single-component upgrades to reduce technology obsolescence that is so common in military space programs. Launch cost reduction has been a US Air Force interest since the 1990s [*Sellers et al.*, 2015], an effort at which SpaceX has been successful so far, as evidenced by their retrieval and reuse of the Falcon 9 rocket booster to reduce recurring launch vehicle costs by up to 70% [*Sheetz*, 2020]. Launch costs have also been reduced by Air Force organizations procuring rideshare opportunities as a hosted payload. Additionally, operating costs that traditionally account for up to 30% of total lifecycle cost [*Sellers et al.*, 2015] can be offset by automation. In summary, being gun-shy with test due to the cost of a space vehicle is no longer an entirely valid reason, and AFRL/RV has been setting the precedent for mitigating risks in the face of expanding capability needs.

However, these cost improvements have not changed the reality on the ground. Testing beyond functionality checkouts in the heart of the envelope is almost non-existent once a spacecraft is placed into orbit. Deficiencies in spacecraft design or behavior are discovered through the course of normal operations, and mitigated by workaround or operator procedure. This is the equivalent of testing an F-16 at 300 knots, delivering it to the warfighter, and when discovering speed instabilities at 310 knots, limiting the aircraft to no greater than 280 knots for the rest of its operational lifespan. It might be a 600-knot aircraft, but one would never know it.

Let us consider a real-world example: the anomaly and recovery of the space vehicle *CloudSat*. In 2011, NASA's pioneering cloud profiling radar satellite *CloudSat* experienced a battery anomaly that placed it a non-mission capable emergency mode. Workarounds to flight software developed by a joint USAF/Jet Propulsion Laboratory team managed to restore sunlit-orbit only operations by cycling the payload and subsystem components in tune with earth eclipse entry in order to maintain positive power and thermal profiles [*Nayak et al.*, 2012]. However, neither the power nor payload was tested in any kind of non-nominal way until the anomaly occurred. For a long time, the recovery of this $225M satellite was in serious doubt, because while the team needed sufficient spacecraft power to evaluate this alternate operations profile, they lacked valuable data on power-deficient functionality of its components. This data could only have been provided by a test

campaign while the battery was fully functional. An easy parallel to the *CloudSat* situation would be testing the ability of a two-engine T-38 to fly on one generator, and restoring both generators after the test was complete; a fairly basic test and one mandatory to complete before declaring the aircraft operational.

AFRL/RV missions have the advantage of satellite developers and engineers on crew and cradle-to-grave involvement, which allows for less conservative issue mitigation. Some flagship NASA missions like *Cassini* have had engineering units on the ground capable of fully emulating the on-orbit asset, but these are the exception. Despite either/both, it is still common for operational missions to overly constrain themselves in the name of safety. Current-day operational space missions focus on mission longevity, survivability and capability above all else, and satellite testing on-orbit is accordingly handcuffed. The paradigm that AFRL/RV has begun to shift must continue if the USSF is to develop into an independent warfighting force.

The space community's unwillingness to explore (let alone push) the boundaries of a vehicle intended for operational capabilities has many consequences. The most critical is lack of knowledge of where the performance boundaries even *are*. There is no envelope expansion plan for new spacecraft; for example, to determine the maximum specific impulse possible for an avoidance maneuver, or the minimum time to desaturate reaction wheels, or the ability of a sensor to image at the most extreme illumination conditions. Instead, the satellite never strays from conservative, requirements-based, heart-of-the-envelope conditions during its on-orbit life. These restrictions (flight rules) become only more severe as complications arise across the course of the mission.

Without truly knowing the vehicle's maximum capabilities, there naturally follows a complete inability for the manufacturer/developer to train the operations crew on what those capabilities are, and a lack of crew knowledge on how to execute them in a time of need. Space testers are the professionals who deliver this capability, but they are generally neither equipped to manage this under the current risk-averse paradigm, nor empowered to use model data to justify test in the TPS-classic "predict-test-validate" approach [*Montes et al., 2018*].

Ultimately this mentality permeates down into daily operations. If the vehicle has not been "cleared" for anything except heart-of-the-envelope operations, any deviation from this norm becomes, in essence, a test event. This is why even the smallest spacecraft failures, such as a transient low-power state that places the vehicle into a pre-designed "safe mode", are briefed up to the executive (O-6 and above) level and require developer input before any action can be taken.

In the space world, there is one phase in which testing vehicle limits is slowly becoming more commonplace: End of Life (EOL) campaigns. In this phase, a space vehicle's mission is coming to a close [*Nayak*, 2013] and it is possible to argue a less protective risk posture with even the most conservative leadership. These higher risk tests provide valuable information about what the space vehicle can do beyond nominal operations; caution is thrown to the wind and the technology is tested to its limits. With the right (proficient) testing cadre, and pre-defined limits and risk mitigation, space test teams can explore more boundaries to understand the capabilities at our disposal, should we need them.

The operational paradox, however, is immediately apparent. The satellite's true envelope is never determined until the very end of its useful life. It must also be mentioned that not all satellites undergo an EOL campaign. Though they provide useful information to spacecraft designers, unless a relevantly similar satellite is launched, EOL capability determinations may be tactically useless, especially from a warfighting perspective. Two of the authors were part of a particular extensive EOL test campaign, only to find that the satellite could have safely taken 35% more images across its lifespan, if the flight rules around imagery criteria had been based on the sensor's true capabilities. This was determined two days prior to burn in over the Pacific Ocean.

While it may seem obvious to those in the air domain, a space test cadre must determine the tactical edges of the satellite's performance envelope at the start of its operational career, not its end. Warfighter-level employment of space capabilities starts with various levels of test, and the fundamental envelope expansion campaign for propulsion, electronic warfare, sensor operations and other systems onboard the spacecraft. This is something that bears relevance to the education of future "long-course" Space TPS graduates.

## 3. Section Four of the Flight Manual: Normal (Permitted) Procedures

A military pilot's trusty (and mandated) traveling companion is the "Dash-1", the technical order (TO) that lists details on systems operation, checklists, normal and emergency procedures. Civilian pilots have an "owner's manual" or pilot's operating handbook (POH) that contains similar information. Both source from the original flight test program, and are intended to be the one-stop reference for day-to-day (Section Four of the flight manual) or contingency (Section Three) operations.

The Section Four for space operators are the flight rules [for an example on the *Cassini* spacecraft, see *Burk and Bates*, 2008; for human spaceflight, see *Barreiro et al.*, 2010]. These generally have strictly-coded sets of permissible behaviors, leaving space operators significantly more constrained than the average air operator. Each space vehicle generally has an extremely limited number of contingency procedures, planned prior to launch and dealing with only the most benign root causes. Essentially, Section Three is limited for space operators; with better test programs this can be addressed. In the case of vehicle emergencies, crews are authorized to bring the vehicle to a safe state, but reporting and approval is required prior to returning the vehicle to a nominal operating (data collecting) mode, unless it is a recurring failure mode that has a documented and understood recovery procedure.

AFRL/RV has made a habit of tactically developing experimental procedures (termed "memograms"), and integrating them into regular contingency procedures or spacecraft automation should they prove successful. Unlike operational (non-experimental?) space squadrons, AFRL/RV is able to accomplish this only because they retain the vehicle I&T team on the operational crew. This should be the goal for the entire space test enterprise as it expands beyond the lab under USSF.

For any space operator, the primary external contingency action at an operator's disposal is the act of declaring a "1-Bravo" (a vehicle emergency) to the Air Force Satellite Control Network (AFSCN). This attracts the unwanted and time-consuming attention of many generals and others in high positions, while merely allowing the crew to secure more contact time on ground-to-space antennas for troubleshooting. Relevant to constellation spacecraft, contingency actions involve effects or recoveries involving one's own vehicles only, such as performing an on-board computer reset or changing a payload mode, as opposed to a more coordinated or tactical constellation-assisted, SA-enhancing recovery.

In order for a space operator to truly have a responsive warfighting capability, space operators need two things: significantly more ambitious capabilities in their Section Three toolbox, and the authorization to use them when the need arises (which may require policy considerations as well). This is not just limited to off-nominal operations. Wartime or engagement contingencies should be written in to operating procedures, and tested during initial deployment of the vehicle. Pilots know what to do if they hit the pickle button and a bomb does not come off the aircraft (hung store contingency). Deployed pilots are cleared for unlimited use of a training laser, and to employ a combat laser if within certain fairly broad limitations defined by AOR. Space operators employing future USSF systems need similarly well-developed contingency and employment plans. A basic example for spacecraft would be the ability to slew through a keep-out angle to point a camera at an enemy, something most systems are not capable of doing without a system shutdown and automatic safing procedure. Space operators should operate in a disciplined manner within the bounds of the Flight Rules (operating limits), but should also know their real-time and riskier options in an emergency/wartime situation.

Beyond that, of course, is the "Dash-34", the manual outlining how operators can employ weapons from the system. As space becomes more contentious, developing tactics, techniques, and training along with policy must evolve to meet these needs. Actions in the satellite Dash-34 toolbox might include the ability to do maneuvers quickly without waiting for a full collision-avoidance check, or using a LIDAR on an intruder vehicle without extensive coordination and approval from the Laser Clearing House. We stress that these are the most basic scenarios; there are many more that will require intensive testing to develop conditions for safe employment, as well as warfighter tactics. The operator must understand the impact of this activity – for example, the possibility of losing fine pointing ability because of a saturated star tracker – and be appropriately trained to weigh the cost and benefit in real-time as a fighter or bomber pilot is today. Similar to declaring a 1-Bravo, employing Dash-34 actions in space could be expected to require cleanup and explanations after the fact, even if operators are authorized to use these in extreme circumstances, maintaining DoD rules of engagement and proper use of force techniques. Well-trained operators are at the crux of enabling this kind of evolution.

### 4. Safety Nets: The Foundation to Safe, Secure, Efficient Flight Test

While "failure testing" pushes the limits of a spacecraft, it does not necessarily present a very high risk. An example in the airplane world is in-air restart testing. Risk to the operator and the SUT are mitigated extensively, through a rigorous safety review board, and technical measures ranging from aircrew training and build-up testing to drive down uncertainty in outcome, to emergency power units or spin chutes or ejection seats. In other words, *safety nets* can be put in place to ensure protection, while refining procedures to deal with expected failures during the course of vehicle operation. These can be from traditional models and simulations, to more tactical, thought-out methods of live implementation.

With satellites, the risk grows because of lack of access to the test asset. A chase spacecraft for added situational awareness is not possible. Telemetry is intermittent, due to limited access periods with ground antennas as the satellite travels around the globe. An onboard failure that could be completely fixable may become an **end-of-life** event if the satellite cannot downlink enough data for the test team on the ground to diagnose the problem. Failure testing, therefore, takes on an added dimension of complexity for the space domain.

AFRL/RV has come to use safety nets regularly in experimental space test, to drive down risk when executing procedures that have uncertain outcomes. Specific to the space domain, one method is to upload time-tagged commands to the vehicle prior to executing higher-risk activities. These will put the vehicle back to a known state, such as changing back to the original guidance mode, shutting off the SUT, or others. If something unexpected occurs, the crew can expect the vehicle to autonomously recover and communicate at a known time. Another methodology is the use of "passively safe" trajectories for Rendezvous and Proximity Operations (RPO) missions [*Breger and How*, 2008]; if the vehicle encounters an issue which makes it unsafe to continue in proximity operations to another space object, guidance can default to off and the vehicle will naturally drift away from the other object. If baked in to the test program and plan from the beginning, similar methods can mitigate the risk posed by executing tests that expand the envelope (see discussion in Section 2). These tests were performed on test/experimental satellites designed with specific capabilities in mind.

Fault response systems on a space vehicle can be both a blessing and a curse, and must be managed carefully by an experienced operator to work as intended. Fault responses generally monitor designated telemetry points, and take a predetermined response action when a point goes out of limits. These responses protect the vehicle from unfavorable conditions, but if the test team intends for the vehicle to be stretched to its limits, it is possible that a triggered fault response may not be desirable. An example of this in the air world is the Automatic Ground Collision Avoidance System (AGCAS). While its positive impact to general aircrew safety is certain [*Koltai et al., 2012*], an unintended AGCAS fly-up while in fingertip formation on an Air Force Thunderbirds aircraft can have disastrous consequences in the name of safety. Such safety systems require special consideration when the vehicle is in a test phase.

The best fault protection system is highly customizable, and allows a proficient operator (discussed further in Section 5) to manage vehicle responses closely. For the space domain, one example would be a $\Delta V$ (fuel) limit set to prevent accidental commanding of an excessive maneuver; in a stressing scenario it may be desirable to execute that aggressive maneuver. It is also critically important that an operator be knowledgeable of *how* their fault detection system works, and what the current settings, are at all times. While obvious, this can be difficult when the mission team is at a fixed location on the Earth, and the satellite is traveling over the horizon from the transmitter within a matter of minutes. We have discussed the need for the test team to be empowered to deal with contingencies appropriately and immediately; the next section covers thoughts on the necessary training level for that team to be successful.

Satellite and ground system developers can aid the case for a more agile, warfighting vehicle by incorporating some of these considerations into vehicle and ground design. For example, redundancy can be added where budgets permit, flight software accommodations can be incorporated for flexible on-board automation and reconfigurable fault responses, and ground systems can be designed for quick-turn modification of complex vehicle activities. All contingency activities should be modeled and simulated, wherever possible. Developers should always consider how their implemented safe modes and fault response structures will impact test actions, contingencies, and responses to stressing or wartime circumstances.

## 5. Who has the con? Space Test Cadre Proficiency and Qualifications

A future space test cadre must be experienced and proficient enough to walk necessary boundaries safely and efficiently, and deliver a system that will enable future operators to defend the warfighting environment of space. Proficiency is a critical pre-requisite. In addition, levels of proficiency or expertise should be employed to single out the most proficient operators for special tasks (the space equivalent of a four-ship flight lead), and to incentivize junior operators to strive for higher levels of proficiency (similar to a proficient wingman upgrading to a two-ship flight lead).

In the air world, each aircraft is its own Mission Design Series (MDS). Within an MDS, pilots obtain a mission qualification based on a series of checkrides and upgrades. Test operators may obtain additional test-specific qualifications such as a chase, airstart, departure or tower fly-by qual. Flight test engineers do the same, both for aircraft, and control room positions such as test directors, conductors, and test essential specialists. Once cleared, operators must then maintain their qualification through a model of task-based currency [*Bigelow et al.*, 2003], i.e., $X$ sorties or $Y$ hours flown every quarter, with test- or aircraft-specific tasks defined and logged for proficiency every $Z$ weeks.

This "reps" based model may not fit experimental (one-of-a-kind) or test (intended for future operational employment) spacecraft, where each vehicle has a unique personality, which can change over the course of the mission life. A space operator cannot depend on recurring objectives to train for; instead, they must know the current state of the vehicle (common quirks and regular housekeeping tasks), the current state of the mission (lockout periods, constraints, frequency of data collection), and the state of procedures (location of updated procedures, changes made and why, and available contingencies). Each vehicle with which an operator has experience will provide crucial data points for dealing with similar situations and subsystems on other vehicles. Additionally, a highly qualified crew member should have "logged" enough practice operating a spacecraft under stress that they have demonstrated an ability to remain calm and diagnostic in any number of demanding, unknown situations.

The key phrase is *under stress*. Once a satellite is placed on orbit, we posit that qualification for high-risk test events should be conducted on an experience model. Since a single misstep could result in the asset being lost, or unrecoverable, *hours in high-risk test per MDS* should become the defining factor for which operators participate in planning and operating a high-risk space test or combat operations. An alternate way of considering it: if a single test could wipe out the entire F-16 fleet, only the most experienced testers would be permitted to execute that test. A schema must exist not only to develop, but to measure and rank operators by experience and competence.

This requires the development of space test training, to give operators valuable experience without actually risking the vehicle. A future "Space Test Pilot School" would have an outsized role to play here. This training can be done first by using simulators, and later by allowing operators-in-training to be shadowed by seasoned operators during high-tempo activities. The seasoned operator can then jump in if things start going awry (backseat IP model).

## 6. Conclusions

The future USSF space test cadre education, proficiency, and safety discussed herein do not start with education, but with a culture shift. The USSF must first normalize a culture shift where risk is warfighting business and where operator proficiency is paramount. Today's paradigm is operators with minimal training on their systems and little authority to make decisions, who rely on checklists provided by the developers and testers, and must call back to their spacecraft provider to solve problems, or to take actions beyond the usual checklist. Tomorrow's paradigm must be to educate and develop Space Test Operators who truly know their systems, and empower them to respond in an appropriate, agile way, as AFRL/RV has begun to do. This calls for upping the ante on space vehicle training, emphasizing operator experience, and joining the currently distinct R&D and organizational space test communities, something that should be part of the discussion when developing the curriculum for a future Space Test Pilot School.

**Acknowledgments.**

**Biographies.**

*Dr. Michael "Orbit" Nayak* is an F-16 and T-7A Flight Test Engineer with the 416th Flight Test Squadron, Air Force Test Center, Edwards AFB, CA, where he leads avionics testing for all blocks of USAF F-16s. He is an FAA-certified CFII, MEI and IGI, and is mission-qualified aircrew in the F-16 and T-38 aircraft. He is a USAF Test Pilot School graduate, NDSEG Fellow, former Lead Flight Director for the experimental spacecraft *TacSat-3* and a member of the Society of Flight Test Engineers.

*Christina "X" Straight* is the founding Operations Cadre Lead for the Space Vehicles Directorate, Air Force Research Laboratory, Kirtland AFB, NM. She first worked experimental satellite operations doing trajectory planning for the *XSS-11* (2005) mission, where she discovered an aptitude for managing intractable satellites and herding ops crews. She has since worked ops for every major AFRL/RV satellite, leading or co-leading mission operations for the last four missions, including serving as a Lead Flight Director for *ANGELS* and the *EAGLE-Mycroft* pair.

*Evelyn "EV" Kent* is the Operations Cadre Deputy for the Space Vehicles Directorate, Air Force Research Laboratory, Kirtland AFB, NM. Her first venture into space operations was on *FalconSat-3* (2009), as a cadet at the US Air Force Academy. She has led or co-led ops development for large and small AFRL satellites, including serving as the Lead Flight Director and Operations Lead for the largest unmanned structure in orbit, *DSX*. Evelyn has logged space vehicle operations time with two FalconSats, a number of National Reconnaissance Office (NRO) vehicles, and four AFRL experimental satellites (and counting).

*Jarred Langhals* is part of the first cadre of Space Force officers, currently at the Space and Missile Systems Center, Innovation and Prototyping Directorate, Kirtland AFB, NM. He has worked space operations for *SHARC*, *EAGLE*, *Mycroft*, *DSX* and *GPIM*, among others.